\documentclass[aps,prb,twocolumn,superscriptaddress,showkeys]{revtex4}
\usepackage{bm,amsmath,graphicx}
\begin{document}

\title{Classical and quantum concepts of probability applied to diffraction physics in perfect crystals}
\author{S\'ergio L. Morelh\~ao}
\email{morelhao@if.usp.br}
\affiliation{Instituto de F\'{\i}sica, Universidade de S\~ao Paulo, CP 66318, 05315-970 S\~aoPaulo, SP, Brazil}
\author{Luis H. Avanci}
\affiliation{Instituto de F\'{\i}sica, Universidade de S\~ao Paulo, CP 66318, 05315-970 S\~aoPaulo, SP, Brazil}

\date{\today}
\begin{abstract}
By following the trajectories of quantum particles inside a periodic lattice and preserving their classical probabilities for reflection, transmission and absorption at each lattice plane, classical scattering outcomes are obtained. Diffraction phenomena in crystals are reproducible after assigning probability amplitudes to every classical outcome. When applied to X-ray diffraction in reflection geometry, this procedure has provided simple recursive equations to calculated the X-ray reflectivity of crystals with thickness varying since a few atomic layers to infinity. The results are in agreement with the dynamical theory of X-ray diffraction, even when absorption is considered.
\end{abstract}

\keywords{Diffraction physics, X-ray photoabsorption, atomic scattering factor}

\maketitle

\section{Introduction}

Diffraction is essentially an undulatory phenomenon and, consequently, diffraction theories were developed using a formalism that describes the propagation of waves through a periodic medium. The propagation of vectorial electromagnetic waves is governed by Maxwell's equations while the propagation of wave functions is governed by Schr\"{o}dinger's equation. However, both type of waves are related to the amplitude of probabilities for possible trajectories of photons or matter particles (electrons, neutrons, ...{\em etc}) inside the medium. Therefore, by calculating the probabilities of all possible trajectories for classical particles, i.e. by visualizing the particles as globules, and assigning to each classical probability quantum probability amplitudes we must be able to achieve a complete quantum description of diffraction phenomena.

This concept of equivalency between classical and quantum probabilities is further developed in this paper and applied to diffraction process in reflection geometry.

\section{General formalism}

The outcome of elastic scattering of classical particles, like bullets, by partially permeable {\em walls} can be calculated by introducing reflection and transmission probabilities. For example, two identical machine guns, symmetrically displaced on each side of a {\em wall}, as shown in Fig.~1, are shooting at the frequency $N_0$. The numbers of bullets per unit of time reaching the D$_1$ and D$_2$ counters are $N_1=(r+\bar{t})N_0$ and $N_2=(\bar{r}+t)N_0$, respectively. $r(\bar{r})$ and $t(\bar{t})$ are reflection and transmission probabilities for bullets from the S$_1$(S$_2$) sources in which $N_1+N_2=2N_0$ since $r+t=1$ and $\bar{r}+\bar{t}=1$.

In the case of identical sources of quantum particles, $\psi_R=R\psi_0$, $\psi_T=T\psi_0$, $\psi_{\bar{R}}=\bar{R}\psi_0$ and $\psi_{\bar{T}}=\bar{T}\psi_0$ are the probability amplitudes assigned to each classical outcome given by the $r$, $\bar{r}$, $t$, and $\bar{t}$ probabilities in Fig.~1. Since $\psi_0$ stands for the amplitude of the incident beams $|\psi_R|^2+|\psi_T|^2=|\psi_{\bar{R}}|^2+|\psi_{\bar{T}}|^2=|\psi_0|^2$, or

\begin{equation}
|R|^2+|T|^2=|\bar{R}|^2+|\bar{T}|^2=1\>.
\label{eq1}
\end{equation}

\begin{figure}
\includegraphics[width=3.2in]{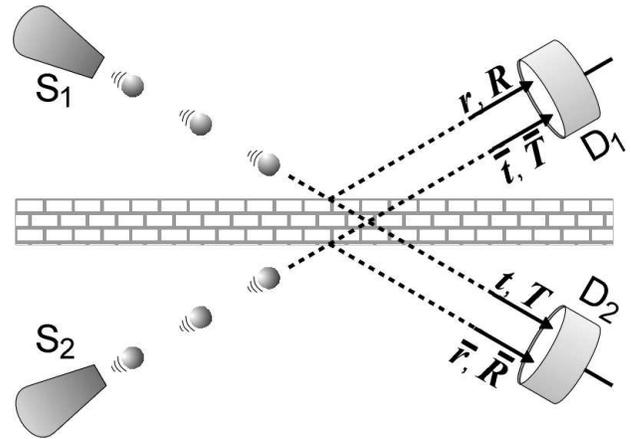}
\caption{Elastic scattering of particles from identical beam sources S$_1$ and S$_2$, symmetrically displaced at both side of a partially-permeable {\em wall}. Classical particles only requires reflection ($r$, $\bar{r}$) and transmission ($t$, $\bar{t}$) probabilities for predicting the outcomes. On the other hand, probability amplitude coefficients, $R$, $\bar{R}$, $T$, and $\bar{T}$ are required for quantum particles such as photons and electrons; their square modulii provide the classical probabilities while their phase relationships obey Eq.~(\ref{eq2}).}
\end{figure}

As in the classical case, the sum of the beam powers (particle energy $\times$ frequency) at both detectors are identical to the total power of the sources. It means that $|\psi_R+\psi_{\bar{T}}|^2+|\psi_{\bar{R}}+\psi_T|^2=2|\psi_0|^2$, and hence 

\begin{equation}
R\bar{T}^{*}+R^{*}\bar{T}+\bar{R}T^{*}+\bar{R}^{*}T=0.
\label{eq2}
\end{equation}
This equation stipulates some phase relationships between the reflected and transmitted amplitudes, as usually obtained for laser beam splitters.~\cite{loudon2000} Without loosing generality, the probability amplitude coefficients can be written as

\begin{eqnarray}
R&=&\pm i\sqrt{r}\>e^{i(\delta+\bar{\varphi})},~~T=\sqrt{t}\>e^{i\varphi},\nonumber\\
\bar{R}&=&\pm i\sqrt{\bar{r}}\>e^{i(\bar{\delta}+\varphi)},~~{\rm and}~~\bar{T}=\sqrt{\bar{r}}\>e^{i\bar{\varphi}}
\label{eq3}
\end{eqnarray}
where according to Eq.~(\ref{eq2}), 

\begin{equation}
|R||\bar{T}|\sin\delta+|\bar{R}||T|\sin\bar{\delta}=0. 
\label{eq4}
\end{equation}
$\varphi$ and $\bar{\varphi}$ stand for phase variation across the thickness of the {\em wall}; their values are arbitrary regarding convenient choices of references. The $\delta$ and $\bar{\delta}$ phases are determined by the internal structure of the {\em wall}, which is in fact a plane of matter with thickness $d$, comparable to the particle's wavelength $\lambda$. When $|R|=|\bar{R}|$, and consequently $|T|=|\bar{T}|$, $\delta=-\bar{\delta}$ according to Eq.~(\ref{eq4}). If $\delta=\bar{\delta}=0$, the reflected waves are $90^{\circ}$ shifted with respect to the transmitted ones.

Although the phase shift of reflected waves is a well-known phenomenon,~\cite{james1950, loudon2000} we have reproduced its demonstration here to emphasize the correlation between classical and quantum probabilities in which the square modulii of the probability amplitude coefficients are related to the classical probabilities as given in Eq.~(\ref{eq3}) where $|R|^2=r$, $|T|^2=t$, $|\bar{R}|^2=\bar{r}$, and $|\bar{T}|^2=\bar{t}$.

This concept can also be applied on a rather complex situation configured by one incident beam of particles and two parallel planes, as depicted in Fig.~2. Since the particles can suffer several bounces (reflections) at both planes before leaving the interplane region, classical reflection and transmission probabilities are given by

\begin{subequations}
\begin{equation}
P_R^{\>class.}=r+tr\bar{t}+tr\bar{r}r\bar{t}+tr\bar{r}r\bar{r}r\bar{t}+...
\label{eq6a}
\end{equation}
and
\begin{equation}
P_T^{\>class.}=t^2+tr\bar{r}t+t r\bar{r}r\bar{r}t+...\>,
\label{eq6b}
\end{equation}
\label{eq6}
\end{subequations}
respectively. They correspond to the sum of probabilities for all possible outcomes of the scattering, in the sense that the particles can be scattered after one, two, ..., or $n$-bounces. Each outcome can be identified as a classical $n$-bounce channel, where channels with odd (even) number of bounces contribute only to the reflection (transmission) probability.

The quantum coefficients of scattering are obtained by assigning to each classical channel, identified in Eqs.~(\ref{eq6}), a probability amplitude with its respective phase. The phase depends on the total length of the trajectory of a given channel. If $\varphi$ and $\bar{\varphi}$ are chosen to be zero in Eq.~(\ref{eq3}), the phase difference between two consecutive channels scattering on a same direction is an explicit function either of the plane thickness $d$, as well as of the separation between them $h$. However, if we choose

\begin{equation}
\varphi = \bar{\varphi}=-\>\frac{2\pi}{\lambda}d\sin\theta\>,
\label{eq7}
\end{equation}
the amplitude coefficients will account for the phase difference owing to the plane thickness. For instance, the probability amplitudes of channels $1^{\prime}$ and $1$ in Fig.~2 are $$\psi_{1^{\prime}}=R\psi_0=\pm i\sqrt{r}\>e^{i(\delta+\bar{\varphi})}\>\psi_0$$ and $$\psi_1=TR\bar{T}\psi_0=\pm i\sqrt{tr\bar{t}}\>e^{i\varphi}e^{i(\delta+\bar{\varphi})}e^{i\bar{\varphi}}e^{i\Delta\varphi(h)}\>\psi_0.$$ Then, $2\varphi+\Delta\varphi(h)$ is the phase difference between two consecutive channels where $\Delta\varphi(h)=-4\pi h\sin\theta/\lambda$. The scattering time delay of higher order channels to with respect to the lower order ones is responsible for the minus sign of this expression, as well as in Eq.~(\ref{eq7}).

\begin{figure}
\includegraphics[width=3.2in]{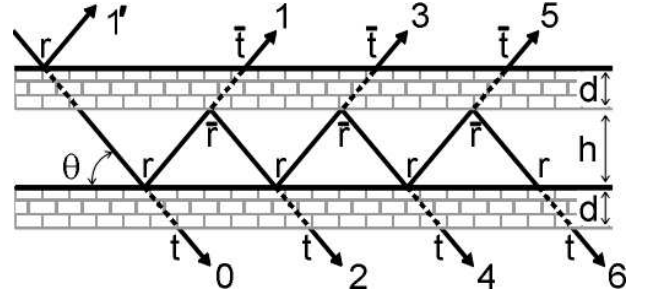}
\caption{Classical $n$-bounce channels for the elastic scattering of particles in a two-plane system. $r,\bar{r}$ and $t,\bar{t}$ are classical probabilities of reflection and transmission for incidence on two different directions: from the top-left or from the bottom-left, e.g. sources S$_1$ and S$_2$ in Fig.~1. $[r,~\bar{r}]+[t,~\bar{t}]=[1,~1]$.}
\end{figure}

Of particular interest is the case where $h\rightarrow0$. In such a case, the amplitude coefficients of two-plane systems are obtained by direct replacements: $r,\>\bar{r},\>t,\>\bar{t}\>\rightarrow\>R,\>\bar{R},\>T,\>\bar{T}$ in Eq.~(\ref{eq6}). It provides $$ R_0 = R[1 + T\bar{T}(1 + \bar{R}R + \bar{R}R\bar{R}R +...)]$$ and $$T_0 = T^2(1 + R\bar{R} + \bar{R}R\bar{R}R+...).$$ for incidence from the top-left, as shown in Fig.~2. In the other symmetrical situation, which corresponds to incidence from the bottom-left, e.g. source S$_2$ in Fig.~1, the $\bar{R}_0$ and $\bar{T}_0$ coefficients for the reflected and transmitted amplitudes are obtained by analogous procedure. Since $$\sum_{n=0}^{\infty}z^n=1/(1-z)$$ if $|z|<1$, the probability amplitude coefficients can be written in a more compact format,

\begin{subequations}
\begin{equation}
\left[\begin{array}{c} 
R_0\\ 
\bar{R}_0
\end{array}\right]=\left(1+\frac{T\bar{T}}{1-\bar{R}R}\right)\left[\begin{array}{c}
R\\
\bar{R}
\end{array}\right] 
\label{eq8a}
\end{equation}
and 
\begin{equation}
\left[\begin{array}{c}
T_0\\ 
\bar{T}_0
\end{array}\right]= \frac{1}{1-\bar{R}R}\left[\begin{array}{c}
T^2\\
\bar{T}^2
\end{array}\right] 
\label{eq8b}
\end{equation}
\label{eq8}
\end{subequations}

The $R_0$, $T_0$, $\bar{R}_0$, and $\bar{T}_0$ coefficients also fulfil the energy/probability conservation equations, Eqs.~(\ref{eq1}) and (\ref{eq2}). Therefore, they provide a building block of recursive formulas for calculating the reflection and transmission amplitude coefficients for N-plane systems, like atomic planes in perfect crystals.

In the present view, each individual scattering plane, with $R$, $\bar{R}$, $T$, $\bar{T}$ coefficients, stands for an element of periodicity composed, in general, of several atomic planes that repeat themselves with period $d$ in order to build up the crystal, as illustrated in Fig.~3. A fast recursive formula to go from very thin to semi-infinity crystals (of infinity thickness) is obtained by building the crystals in geometrical progression of N$=2^n$ planes ($n = 1, 2,...\>$), and hence

\begin{subequations}
\begin{equation}
\left[\begin{array}{c} 
R_n\\ 
\bar{R}_n
\end{array}\right]=
\left(1+\frac{T_{n-1}\bar{T}_{n-1}}{1-\bar{R}_{n-1}R_{n-1}}\right)\left[\begin{array}{c}
R_{n-1}\\
\bar{R}_{n-1}
\end{array}\right] 
\label{eq9a}
\end{equation}
and
\begin{equation}
\left[\begin{array}{c} 
T_n\\ 
\bar{T}_n
\end{array}\right]= \frac{1}{1-R_{n-1}\bar{R}_{n-1}}\left[\begin{array}{c}
T_{n-1}^2\\
\bar{T}_{n-1}^2
\end{array}\right] 
\label{eq9b}
\end{equation}
\label{eq9}
\end{subequations}
are the amplitude coefficients for a crystal of thickness $\tau = {\rm N}d$. It is obtained by analogy with the coefficients of the two-plane system in Eqs.~(\ref{eq8}).

The maximum of the intensity reflectivity $|R_n(\theta)|^2$, occurs around the diffraction condition of the lattice, where $\varphi = -m\pi$ and $m$ is an integer number. The center of the first order reflection, $m=1$, is observed at the Bragg angle $\theta_B$.

\subsection{Absorption}

There are two methods to properly allow absorption in the medium. They differ on the distribution of the interplane matter~\textendash~the matter that fills up the space between the ideal planes used to represent the periodicity of the structure.

\subsubsection{Method I}

Depending on the interaction between the particles and the atoms in the crystal, some of the particles will not be able to further participate on the diffraction process after reaching an atomic plane. It gives rise to general absorption probability, $a$, accounting for the particles that are not reflected neither transmitted under diffraction condition, so that $t+r=1-a$. 

Since $r$ and $a$ are very small quantities compared to $t\simeq1$, the reduction in the reflection probability $r(1-a)\simeq r$, due to absorption is negligible in comparison to the reduction in the transmission probability $t(1-a)\simeq t-a$. Consequently, absorption is taken to reduce the transmission probability only, i.e. $t=1-r-a$ in Eq.~(\ref{eq3}). 

In this method, all diffraction channels of the two-plane system in Fig.~2, except channels $1^{\prime}$, have the same absorption probability of $2a$. It corresponds to the intensity reduction ratio of $a$ per interplane distance, and hence to an effective linear absorption coefficient $\mu = a\sin\theta/d$. It certainly is a valid approach when the interplane space is empty; for instance, when the elements of periodicity are made of single atomic planes.

\begin{figure}
\includegraphics[width=3.2in]{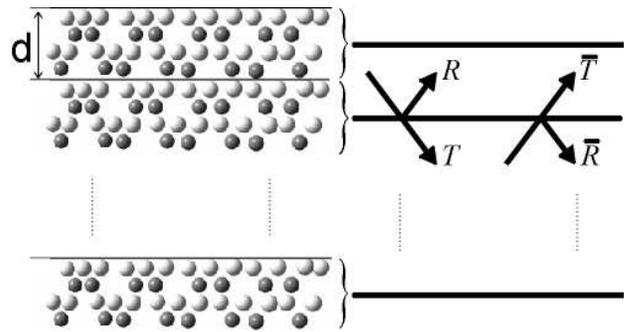}
\caption{Elements of periodicity in crystals are composed, in general, of several atomic planes that repeat themselves with period $d$. The scattering properties of each element is given by reflection $R,\bar{R}$ and transmission $T,\bar{T}$ coefficients, Eq.~(\ref{eq3}). The intensity reflectivity of an array with N$=2^n$ repetitions can be calculated by Eqs.~(\ref{eq9}).}
\end{figure}

\subsubsection{Method II}

When the interplane space is filled up with matter of linear absorption coefficient $\mu$, the absorption of a single plane can still be represented by the absorption probability 

\begin{equation}
a\simeq\mu d/\sin\theta.
\label{eq9.2}
\end{equation}
But, it does not account for absorption along of the zigzag routes of the channels (Fig.~2). Every time the particle crosses the interplane distance, its phase is delayed by $\varphi$ and its exiting probability (from the interplane region) decreased by $\exp(-\mu d/\sin\theta)$. Then, the replacements of $\varphi$ and $\bar{\varphi}$ in Eq.~(\ref{eq3}) by

\begin{equation}
\varphi^{\prime}= -\frac{2\pi}{\lambda}d\sin\theta + i\frac{d}{2\sin\theta}\mu
\label{eq10}
\end{equation}
provide the proper absorption in crystals. 

Since the ideal planes are responsible for the elastic scattering process only, the reflection and transmission probabilities are preserved in this method, i.e. $t+r=1$ in Eq.~(\ref{eq3}). Absorption is taken to reduce the square modulii of the coefficients so that $$|R|^2=re^{-a}\simeq r~~{\rm and}~~|T|^2=te^{-a}\simeq t-a$$ where $e^{-a}\simeq1-a$. The same occurs to $\bar{R}$ and $\bar{T}$. These relationships demonstrate that, the elastic scattering of each element of periodicity is not affected by absorption and, although, methods I and II seems to be physically different, they are in fact equivalent and provide the same results for $a\ll1$. Both methods can be used, depending on particular conveniences, but they must not be mixed. In more specific words, for a same absorption process, one must use $t=1-r-a$ and $\varphi$ [Eq.~(\ref{eq7})], or $t=1-r$ and $\varphi^{\prime}$ [Eq.~(\ref{eq10})].

A few examples on the behavior of the intensity reflectivity, $|R_n(\theta)|^2$, as a function of incidence angle, crystal thickness, reflection and absorption probabilities are shown in Fig.~4. No difference has been observed when using methods I or II for absorption. The maximum reflectivity is always less or equal than 1 even for semi-infinity crystals. The transmitted intensities are given by $|T_n(\theta)|^2$, which is quite different than $[1-|R_n(\theta)|^2]e^{-\alpha\tau}$ in the center of the diffraction peak, as can be seen in Fig.~4 (top-left inset).  The phase of $R_n(\theta)$ is also shown; $\delta=\bar{\delta}=0$ in all cases. The analysis of the reflectivity profiles for thick low-absorbing crystals leads to an empirical equation for the full width of the half maximum (FWHM),

\begin{equation}
W = \frac{2}{3}\sqrt{|R||\bar{R}|}\tan\theta_B.
\label{eq11}
\end{equation}

\subsection{Single bounce crystals}

As a final example of classical-quantum concepts of probabilities applied to diffraction physics, let consider a special crystal where back-reflections are not allowed, i.e. where $\bar{r}=0$ and so $\bar{t}=1-a$ (absorption according method I). In this crystal diffraction occurs by interference of probability amplitudes for a single scattering event in one of N planes.

The classical probability is provided by $$P_{sb}^{\>class.} = r + tr\bar{t} + ttr\bar{t}\bar{t} +... = r\sum_{n=0}^{\rm N-1}(t\bar{t})^n=r\frac{1-(t\bar{t})^{\rm N}}{1-(t\bar{t})}$$ for $t+r=1-a$. However, in this case, for applying the same rule used before to convert a classical problem into its quantum counterpart, as from Eqs.~(\ref{eq6}) to Eqs.~(\ref{eq8}); the probability amplitude of the incoming particle has to be extended to the entire lattice.

To better visualize this, imagine the following situation, N position detectors of area $A_n$ distributed, without overlapping, over a screen of area $A_S$. If a quantum particle has the same probability to hit any position of the screen, the probability amplitude at the $nth$-detector is $\psi_n=\sqrt{A_n/A_S}$ so that $$\sum_{n=0}^{\rm N-1}|\psi_n|^2 = \frac{1}{A_S}\sum_{n=0}^{\rm N-1}A_n$$ is the effective probability of measuring the particle in one of the detectors.

By analogy, the amplitude reflection coefficient, $R_{sb}$, for a single bounce in one of N planes has to be normalized by $\sqrt{N_S}$, where $N_S=\sqrt{N_{\downarrow}N_{\uparrow}}$ is the potential number of planes that the particle can reach on its way down and up inside the crystal. The reflection coefficient is then obtained from $P_{sb}^{\>class.}$ as

\begin{equation}
R_{sb}=\frac{R}{\sqrt{N_S}}\frac{1-(T\bar{T})^{\rm N}}{1-T\bar{T}}.
\label{eq12}
\end{equation}
$R$, $T$ and $\bar{T}$ are given in Eq.~(\ref{eq3}). For transmission probabilities $t$ and $\bar{t}$, the particle can reach a maximum of $N_{\downarrow} = 1/(1-t)$ and $N_{\uparrow} = 1/(1-\bar{t})$ planes, respectively. With this normalization, the maximum possible value of $|R_{sb}|^2$ is 1, for $0<a\ll1$, independently of the real number of N planes in the crystal as shown in Fig.~5.

\begin{figure}
\includegraphics[width=3.2in]{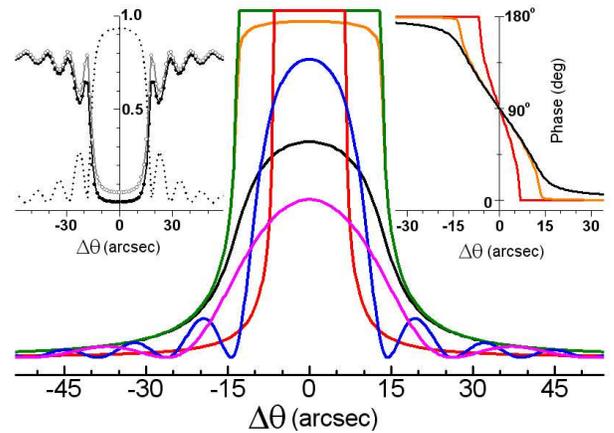}
\caption{Intensity reflectivity $|R_n(\theta)|^2$, calculated according Eqs.~(\ref{eq9}) in crystals of N$=2^n$ planes. $\Delta\theta=\theta-\theta_B$, $\lambda=1.54$\AA, and $d=3.14$\AA~in all of the following cases: (1) [N, $r$, $\mu (cm^{-1})$] = [2048, $16\times10^{-8}$, $0.01$] (pink curve); (2) [4096, $16\times10^{-8}$, $0.01$] (blue curve); (3) [$\infty$, $16\times10^{-8}$, $0.01$] (red curve); (4) [$\infty$, $64\times10^{-8}$, $0.01$] (green curve); (5) [$\infty$, $64\times10^{-8}$, $200$] (orange curve); (6) [$\infty$, $64\times10^{-8}$, $8000$] (black curve); and (7) [4096, $64\times10^{-8}$, $400$] (dashed curve, top-left inset). $|T_n(\theta)|^2$ (closed-black circles) and $[1-|R_n(\theta)|^2]e^{-\alpha\tau}$ (open-gray circles) are compared in the top-left inset for case (7). For cases (3), (4), (5) and (6), the phases of $R_n(\theta)$ are shown in the inset at the top-right.}
\end{figure}

\section{Discussion on X-ray diffraction}

Diffraction of X-rays in crystals is the most common technique for studying how atoms and molecules organize themselves in the solid matter. To apply or compare the intensity reflectivity obtained from Eqs.~(\ref{eq9}) with X-ray results, scattering coefficients for the amplitude of the electric field are required.

The scattering and absorption of X-rays by individual atoms can be described by

\begin{subequations}
\begin{equation}
\sigma_s(2\theta) = r_e \lambda[f(2\theta)+f^{\prime}]e^{-B(2\theta)}
\label{eq12.2a}
\end{equation}
and
\begin{equation}
\sigma_a = 2 r_e \lambda f^{\prime\prime}
\label{eq12.2b}
\end{equation}
\label{eq12.2}
\end{subequations}
when the X-ray polarization is linear and perpendicular to the scattering plane. The former provides the amplitude of the atomic elastic scattering with corrections for atomic resonance $f^{\prime}$, and Debye temperature factor $B(2\theta)$. $f(2\theta)$ is the scattering factor through an angle $2\theta$ from the incident beam direction, and $r_e=2.818\times10^{-5}$\AA~is the classical electron radius. The atomic absorption cross section $\sigma_a$, depends only on $f^{\prime\prime}$ correction of the atomic scattering factor for absorption. More details on $f^{\prime}$ and $f^{\prime\prime}$ are available on the {\em International Tables for Crystallography}.

A single layer of atoms, all of the same kind, with $M$ atoms per unit area scatters an incident plane wave of amplitude $E_0$ according to 

\begin{figure}
\includegraphics[width=3.2in]{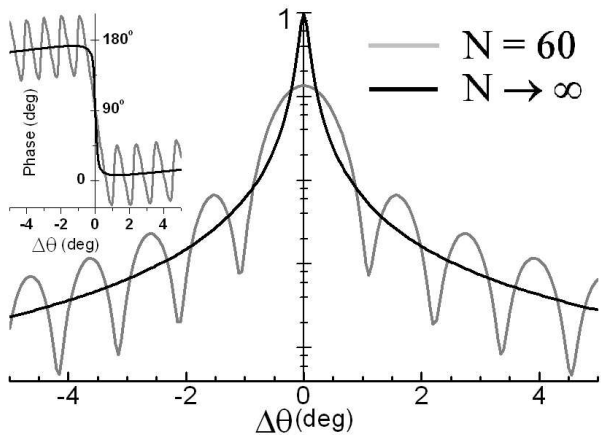}
\caption{Intensity reflectivity $|R_{sb}|^2$, in {\em single bounce} crystals according to Eq.~(\ref{eq11}). $\Delta\theta=\theta-\theta_B$, $\lambda=1.5$\AA, $d=1.0$\AA, $r=0.01$, $a=0.002$, and ${\rm N}=60$ (gray curve) or ${\rm N}=10^6$ (black curve). The phase of $R_{sb}$ is shown in the inset.}
\end{figure}

\begin{equation} 
E_{SL} = -i\frac{M}{\sin\theta}\sigma_s(2\theta) E_0 = R_{SL} E_0 
\label{eq13}
\end{equation} 
(Warren, 1969). The probability to measure the specular scattering of a single photon by this layer of atoms is, therefore, $r = |R_{SL}|^2$ while the transmission probability is given by $t=1-r-a$ (absorption method I, \S II.A.1.) where

\begin{equation} 
a = \frac{M}{\sin\theta}\sigma_a 
\label{eq13.2}
\end{equation} 
is the absorption probability per atomic layer.

It is possible to generalize the scattered amplitude of the single layer $ E_{SL}$, to elements of periodicity composed of several atomic planes by using the structure factor $F_{\rm H}$, of the chosen reflection H. In symmetrical reflection geometry, the diffraction vector $\bm{H}$, of reflection H, is aligned to the crystal surface normal direction, and hence $$M \sigma_s(2\theta) \rightarrow \frac{r_e \lambda d}{V_c}F_{\rm H}\>;$$ $V_c$ is the unit cell volume and $d = 1/|\bm{H}|$ as usual. With this substitution in Eq.~(\ref{eq13}), the reflection coefficients are obtained as

\begin{equation}
\left[\begin{array}{c} 
R\\ 
\bar{R}
\end{array}\right]=-i\frac{r_e\lambda|C|d}{V_c\sin\theta}\left[\begin{array}{c}
F_{\rm H}e^{i\varphi}\\
F_{\bar{\rm H}}e^{i\bar{\varphi}}
\end{array}\right] 
\label{eq14}
\end{equation}
where $|C|=1$ or $\cos2\theta$ for $\sigma$ or $\pi$ polarizations, respectively. Note that in this development $F_{\rm H}$ is accounting for the elastic scattering only. It means that,

\begin{equation}
F_H = \sum_n (f + f^{\prime})_n\exp(2\pi i\bm{H}\cdot\bm{r}_n).
\label{eq14.2}
\end{equation}
$n$ runs over all atoms in the unit cell, and temperature factors are now implicit in the $f$ and $f^{\prime}$ values. 

Within the approximation of uniform absorption by the interplane matter, the linear absorption coefficient 

\begin{equation}
\mu = \frac{2\pi}{\lambda}\Gamma\sum_n f_n^{\prime\prime}
\label{eq14.4}
\end{equation}
where $\Gamma = r_e\lambda^2/\pi V_c$, is obtained from the average absorption probability per element of periodicity in which $$M \sigma_a \rightarrow \frac{2 r_e \lambda d}{V_c}\sum_n f_n^{\prime\prime}$$ in Eq.~ (\ref{eq13.2}); then Eq~(\ref{eq9.2}) provides the above $\mu$ expression. 

Since $R$ and $\bar{R}$ are known, it is possible to compare the intensity reflectivity from Eqs.~(\ref{eq9}) with that expected from the dynamical theory of X-ray diffraction~\cite{ewald1916, ewald1917, laue1930, james1950, zacha1945, authier1961, batter1964a} in symmetrical reflection geometry.~\cite{darwin1914a, darwin1914b, prins1930} The width of the total reflection domain, characterized by top-hat shape of the reflection curves, as in Fig.~4, also known as the Darwin width is, according to Authier \& Malgrange,~\cite{authier1998}

\begin{equation}
W_D = \frac{2d}{\pi}\frac{r_e\lambda|C|(F_{\rm H} F_{\bar{\rm H}})^{1/2}}{V_c \cos\theta}=\frac{3}{\pi}W.
\label{eq15}
\end{equation}

By replacing Eq.~(\ref{eq14}) into Eq.~(\ref{eq11}), the ratio between $W_D$ and the width $W$, of the $|R_n(\theta)|^2$ curve, is observed to be very close to unit. The phase behavior of $R_n(\theta)$, inset Fig.~4, is also very similar to that calculated by the dynamical theory~\cite{authier1986} although the center of the reflection curve at $\theta_B$ are slightly different when $\lambda$ is corrected for the effect of refraction.

The intensity reflectivity calculated here provides similar results to the dynamical theory. The developed recursive formula, Eqs.~(\ref{eq9}), is quite equivalent to that developed by Darwin;~\cite{darwin1914a, darwin1914b} although the actual deduction has been obtained by a simple procedure. It is almost as simple as that of the Laue's kinematical (or geometrical) theory,~\cite{fried1913, zacha1945, warren1969} which was derived by adding the amplitudes of the waves scattered by each plane, and taking into account only the optical path differences among them; but neglecting the interaction of the propagating waves and matter. It is possible to derive the kinematical reflection coefficient from $R_{sb}$, Eq.~(\ref{eq12}), when $t=\bar{t}=1$ and $N_S=N$, which means that all planes contribute evenly, independently of how deep they are inside the crystal.

\subsection{On the physical meaning of a complex atomic scattering factor}

The seminal work of Darwin~\cite{darwin1914a, darwin1914b} followed by Prins~\cite{prins1930} gave rise to the actual method in which X-ray photoabsorption is accounted in crystals. It is responsible for the $f''$ correction and to the widely used expression of the atomic scattering factor $\tilde{f}=f+f'+if''$, as a complex number.

When calculating the index of refraction for an aggregate of atoms, the imaginary part of $\tilde{f}$ does have a very clear meaning, which is to reduce the amplitude of the traveling waves in the medium due to absorption, as in absorption method II (\S II.A.2.). However, what would be its meaning when used, for instance, to calculate the scattered amplitude by the layer of atoms in Eq.~(\ref{eq13})?

$R_{SL}=-ir_e\lambda M \tilde{f}/\sin\theta$ would lead to $$r=|R_{SL}|^2=\left(\frac{r_e\lambda M}{\sin\theta}\right)^2[(f+f')^2+(f'')^2],$$ which implies that the reflection probability increases as the absorption cross section increases since the latter is proportional to $f''$. This increase in $r$ owning to absorption has no physical meaning; although in practice, this increase is very small $(f'')^2\ll(f+f')^2$, the sum of the reflection and transmission probabilities is preserved $t+r=1$, and the overall result is very close to that obtained by absorption method II.

As a consequence of calculating reflection coefficients only with $f$ and $f'$, the structure factors in Eqs.~(\ref{eq14}) and (\ref{eq14.2}) do not break Friedel's Law, i.e. $|F_H|=|F_{\bar{H}}|$ even in non-centrosymmetric crystals. Moreover, in the present description of the X-ray diffraction, based on preservation of the reflection, transmission and absorption probabilities of X-ray photons, there is still another problem regarding the index of refraction. The atomic absorption cross section $\sigma_a$, Eq.~(\ref{eq12.2b}), is so small that the arrangement of the atoms inside the unit cell does not affect the absorption probability of the X-ray photons when crossing the elements of periodicity. Therefore, the only contribution of $f''$ to be considered in the index of refraction is the average one that provides the linear absorption coefficient in Eq.~(\ref{eq14.4}). It implies that the absorption probabilities along the incident and reflected beam directions have the same value, and hence $|T|=|\bar{T}|$. 

All of the above facts are in agreement with the beam splitter equation Eq.~(\ref{eq4}), where $|R||\bar{T}|=|\bar{R}||T|$ and the structure factor phases have opposite signals $\delta=-\bar{\delta}$. Nevertheless, the intensity reflectivities of the H and $\bar{\rm H}$ reflections can be slightly different. This difference has its root on the imaginary part of $\tilde{f}$, which is the $f''$ correction for atomic photoabsorption, as well as on the formation of standing waves~\cite{borr1941, laue1949, borr1950} inside the crystal.

In reflection geometry, the pattern of standing waves formed by the interference of the incident and reflected waves was proposed by Batterman~\cite{batter1964b, batter1969} and it is a well-known phenomenon today.~\cite{bedzyk1988, patel1996} In the present context, the nodes and antinodes of the standing waves determine the positions where the photons have, respectively, the highest and the lowest probabilities to be found along their classical path inside the crystal. The absorption probability is, therefore, weighted by the pattern of standing waves.

As the crystal is rocked through the reflection domain, the phase of the reflection coefficient varies from $180^{\circ}+\delta$ to $\delta$, as shown in Fig.~4 for $\delta=0$. This phase variation induces the pattern of nodes and antinodes to grid by half a lattice plane distance inside the crystal, and then, the photons have higher probability to be absorbed at only one half of the element of periodicity. For the $H$ and $\bar{\rm H}$ reflections, the one halves scanned by the nodes of the standing waves are not the same and, consequently, the effective absorption for the Friedel's pair can be different. This hypothesis is illustrated in Fig.~6.

\begin{figure}
\includegraphics[width=3.2in]{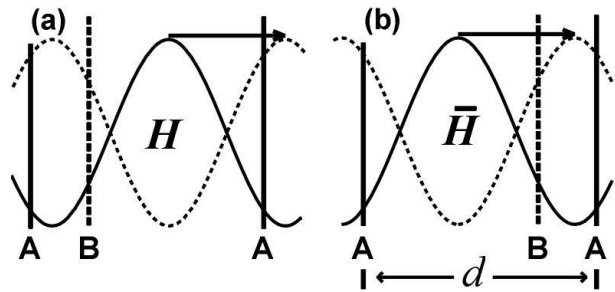}
\caption{Griding of standing waves during rocking curves. The element of periodicity for the chosen reflection H is composed by the A and B atomic planes. The nodes of the standing waves are (a) passing over plane A for reflection H, and (b) over plane B for reflection $\bar{\rm H}$. The arrows indicates the movement of the nodes as the crystal is rocked from $\Delta\theta = -W_D/2$ (solid lines) to $\Delta\theta = +W_D/2$ (dashed lines).}
\end{figure}

To estimate the effective absorption coefficient $\mu_{eff}$ during a rocking curve, the intensities of the standing waves as a function of depth $z$,~\cite{authier1998} are approximated to

\begin{eqnarray}
I_{SW}(z) & \simeq & \frac{I(z)}{2}\left[1+\cos\left(\frac{2\pi z}{d}+\Psi\right)\right]\nonumber
\\
& \simeq & I_0(z)\cos^2\left(\frac{\pi z}{d}+\frac{\Psi}{2}\right) 
\label{eq16}
\end{eqnarray}
where $\Psi$ is the phase of the reflection coefficient $R_n$, accounting for the $180^{\circ}$ phase shift (Fig.~4, inset) across the Darwin width plus the phase $\delta$ of $F_H$. It has also been assumed that the crystal is thick enough to assure a comparable strength between the incident and reflected waves at all depth. Since $I(z)$ is a smooth function with very small variation over the lattice period,

\begin{equation} 
\mu_{eff}(\Psi) = \frac{2\pi}{\lambda}\Gamma\sum_n f_n^{\prime\prime}cos^2(\pi\bm{H}\cdot\bm{r}_n+\Psi/2).
\label{eq17} 
\end{equation} 
This equation is obtained from Eq.~(\ref{eq14.4}) when using the intensity variation of the standing waves, $cos^2(\pi\bm{H}\cdot\bm{r}_n+\Psi/2)$, as a weight function for the atomic absorption cross sections at each $\Psi$ value. $\bm{H}\cdot\bm{r}_n$ stands for the depth of the {\em nth}-atom along the interplane distance.

\begin{figure}
\includegraphics[width=3.2in]{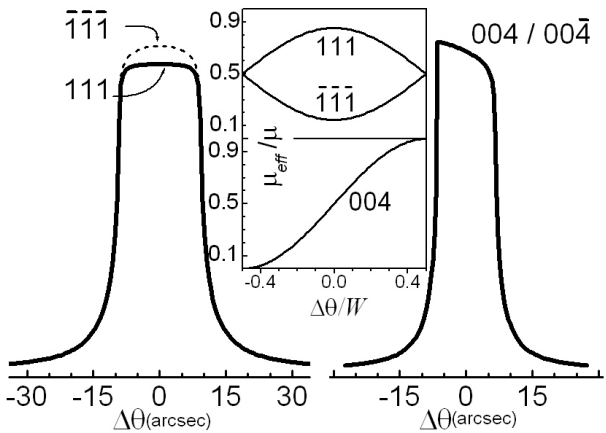}
\caption{Intensity reflectivity accounting for absorption modulation due to standing waves inside a Ge crystal. $\lambda = 1.54$\AA, $f+f'=30.9$, $f''=0.92$, and $\mu=352cm^{-1}$. From $\Delta\theta=-W_D/2$ to $+W_D/2$, the intensity curves are calculated by using $\mu_{eff}(\Psi)$, Eq.~(\ref{eq17}), instead of $\mu$ in Eq.~(\ref{eq10}). The effective absorption coefficient through the Darwin width is shown in the inset.}
\end{figure}

During a rocking curve, the average effective absorption due to modulation by standing waves is estimated by integrating $\mu_{eff}(\Psi)$ in the reflection domain, $$<\mu_{eff}>=\frac{1}{\pi}\int_{\delta}^{\pi+\delta}\mu_{eff}(\Psi){\rm d}\Psi.$$ By solving the integral we have that

\begin{equation}
<\mu_{eff}> = \frac{\mu}{2}-\frac{2}{\lambda}\Gamma\sum_n
f_n^{\prime\prime}\sin(2\pi\bm{H}\cdot\bm{r}_n+\delta). 
\label{eq18}
\end{equation}

Note that the sum of $<\mu_{eff}>$ and $<\bar{\mu}_{eff}>$, for reflection $\bar{\rm H}$, is equal to the linear absorption coefficient $\mu$. For instance, in a germanium crystal and $\lambda=1.54$\AA, $\mu = 352 cm^{-1}$ while $<\mu_{eff}>=255cm^{-1}$ and $<\bar{\mu}_{eff}>=97 cm^{-1}$ for the $111$ and $\bar{1}\bar{1}\bar{1}$ reflections, respectively. Since the average absorptions are not the same, the intensities of the H and $\bar{\rm H}$ reflections can be different.

However, the variation of $\mu_{eff}(\Psi)$ across the reflection domain also changes the profile of the intensity reflectivity curves as demonstrated for a few examples in Fig.~7. A detailed review on X-ray absorption and standing waves in reflection geometry can be found elsewhere.~\cite{batter1964a}

\section{Concluding remarks}

The theoretical foundations of diffraction physics were settled since the beginning of the 20th century. At that time the quantum of light was just a concept proposed by Einstein to explain photoelectric effect, and it was not experimentally proved before 1925, in the same year that Heisenberg developed the basis of the quantum mechanics. Therefore, the Maxwell's electromagnetic theory was the only possible treatment for the X-ray diffraction in the early 1900s. Ever since, a huge number of articles and books have further developed and reported the seeding theories of X-ray diffraction as essentially an undulatory phenomenon. Diffraction of matter particles like electrons and neutrons has received a similar treatment although by the Schr\"{o}dinger's wave equation.

In nowadays it is well accepted that in the helm of quantum reality probability amplitudes rule the existence of the particles of light or of matter. The most fundamental distinction between classical and quantum phenomena are the interference of probability amplitudes that do not exist in the former. By following the trajectories of quantum particles inside a periodic lattice and preserving their classical probabilities for reflection, transmission and absorption at each element of periodicity, series of classical outcomes are obtained. Diffraction in crystals are reproducible after assigning, according to some previously established criteria, probability amplitudes to every classical outcome. When applied to X-ray diffraction in reflection geometry, this procedure has provided simple recursive equations to calculated the X-ray reflectivity of crystals with thickness varying since a few atomic layers to infinity. The results are in perfect agreement with the dynamical theory of X-ray diffraction, even when absorption is considered.  It offers an alternative description of the diffraction phenomenon, with a different point-of-view of the usual wave diffraction theories.

\begin{acknowledgments}
The authors would like to thanks Dr. M\'{a}rcia Fantini for valuable discussions, as well as for the very kind revision of the manuscript.  This work was supported by the Brazilian founding agencies FAPESP, grant number 02/10387-5, and CNPq, proc. number 301617/95-3.
\end{acknowledgments}

\end{document}